\documentclass[prl,twocolumn,showpacs,superscriptaddress]{revtex4}

\usepackage{graphicx}

\usepackage{ulem}
\usepackage{color}
\usepackage{amssymb}

\newcommand{\Ns}{\ensuremath{N_s}}
\newcommand{\Ne}{\ensuremath{N_e}}

\newcommand{\etal}{\textit{et al.}}
\newcommand{\redsout}[1]{}

\begin{document}
\title{Soft Hubbard gaps in disordered itinerant models with short-range interaction}
\author{Hiroshi Shinaoka}
\affiliation{Department of Applied Physics, University of Tokyo, 7-3-1 Hongo, Bunkyo-ku, Tokyo 113-8656, Japan}
\author{Masatoshi Imada} 
\affiliation{Department of Applied Physics, University of Tokyo, 7-3-1 Hongo, Bunkyo-ku, Tokyo 113-8656, Japan}
\affiliation{CREST, JST, 7-3-1 Hongo, Bunkyo-ku, Tokyo 113-8656, Japan}
\date{\today}
\begin{abstract}
We study the Anderson-Hubbard model in the Hartree-Fock approximation and the exact diagonalization under the coexistence of short-range interaction and diagonal disorder. We show that there exist unconventional soft gaps, where the single-particle (SP) density of states (DOS) $A$ follows a scaling in energy $E$ as $A(E)\propto \exp[-(-\gamma\log |E-E_F|)^d]$ irrespective of electron filling and long-range order. Here, $d$ is the spatial dimension, $E_F$ the Fermi energy and $\gamma$ a non-universal constant. We propose a multi-valley energy landscape as their origin. Possible experiments to verify the present theory are proposed.
\end{abstract}


\maketitle

Metal-insulator transitions (MIT) have been a fundamental issue in condensed matter physics for a long time. The MIT is driven either by electron correlations, for example, as Mott transitions~\cite{mott49}, or by random potentials as Anderson transitions~\cite{anderson58}. When the interaction causes an insulator, it opens a SP gap at the Fermi energy $E_F$. The Mott gap and a gap induced by an antiferromagnetic order (AF) are typical examples.  On the other hand, the Anderson insulator exhibits absence of the gap with nonzero DOS at $E_F$, where the insulators are characterized not by the vanishing carrier number but by a singular relaxation time. This makes fundamental differences in low-energy excitations between the Anderson and Mott insulators.

In real materials, however, electron correlations and randomness inevitably coexist, which may take on aspects qualitatively different from the simple Anderson or Mott insulators~\cite{criticality,Kravchenko95}. In particular, under the influence of the interaction, Anderson insulators show qualitatively different feature. Efros and Shklovskii~\cite{Efros-Shklovskii} (ES) have clarified that in the Anderson insulator with the long-range Coulomb interaction, a soft Coulomb gap opens in the SP DOS, $A(E)$ with a power-law scaling as $A(E) \propto {| E-E_F|}^{\alpha}$, $\alpha=d-1$ near $E_F$. The validity of the ES theory was confirmed numerically and in experiments~\cite{ES-book} later. In contrast, within the ES theory, short-range interactions do not generate soft gaps. 

Even for short-range interaction, however, soft gaps with $\alpha\simeq 0.5$ were reported in a Hartree-Fock (HF) study in three dimensions (3D)~\cite{Fazileh06}. Recent numerical studies in two dimensions also show the suppression of DOS near $E_F$~\cite{Chiesa08, Song08}. These suggest the presence of an unconventional mechanism which suppresses the DOS even with short-range interaction. In contrast, a numerical study with the dynamical mean-field theory (DMFT) claimed nonzero $A(E_F)$ even in the insulating phases~\cite{Dobrosavljevic97}. Several mean-field studies gave similar results~\cite{Dobrosavljevic03, Byczuk05}. We clearly need further studies for comprehensive understanding of the short-range case.

Since the dielectric constant diverges at the MIT, effects of the long-range part of the Coulomb interaction are restricted to low energies and the short-range part dominates electronic structures in the experimental energy scale near the MIT. Therefore, unconventional soft gaps, if they exist, may be observed near MITs in real measurements. Indeed, recent photoemission results of SrRu$_{1-x}$Ti$_x$O$_3$~\cite{Kim06, Maiti07} as well as a HF study with the long-range Coulomb interaction~\cite{Epperlein97} indicate breakdown of the ES scaling near the MIT in 3D.

In this letter, through numerical analyses of DOS at energies lower than those of the previous studies, we show even short-range interaction drives opening of a soft gap irrespective of the electron filling, originating from a mechanism entirely different from the ES theory. We call it \textit{soft Hubbard gap}. We show numerical evidences of the soft Hubbard gaps with the HF approximation in one and three dimensions, where DOS $A(E)$ follows an unconventional scaling in energy $E$ as $A(E)\propto \exp[-(-\gamma\log |E-E_F|)^d]$ with $\gamma$ being a non-universal constant. Further support by the exact diagonalization (ED) in one dimension (1D) is given. This scaling reduces to a power-law decay of $A(E)$ toward $E_F$ for $d=1$ and even a faster decay for $d>1$ in contrast to the previous HF study~\cite{Fazileh06}. To clarify the origin of the soft gap, we propose a phenomenological theory. The phenomenology is further numerically tested in detail against the HF results in 1D.

The Anderson-Hubbard Hamiltonian is defined by
\begin{equation}
\mathcal{H}=-t\sum_{\langle i,j \rangle,\sigma}c_{i\sigma}^{\dagger}c_{j\sigma} + U \sum_{i} n_{i \uparrow}n_{i \downarrow}+ \sum_{i, \sigma} (V_{i}-\mu) n_{i\sigma},
\end{equation}
on lattices with $\Ns$ sites and $\Ne$ electrons, where $t$ is a hopping integral, $U$ the on-site repulsion, $c_{i\sigma}^\dagger$ ($c_{i\sigma}$) the creation (annihilation) operator for an electron with spin $\sigma$ on the site $i$, $n_{i\sigma}=c_{i\sigma}^\dagger c_{i\sigma}$ and $\mu$ the chemical potential. The random potential $V_i$ is spatially uncorrelated and assumed to follow two models of the distribution $P_V(V_i)$: the box type of width $2W$, $P_V(V_i)=1/2W$ ($|V_i|<W$) with the average $\langle V_i \rangle =0$, and the Gaussian type, $P_V(V_i) =\frac{1}{\sqrt{2\pi} \sigma} \exp( -{V_i^2}/{2\sigma^2})$ ($\sigma^2 = W^2/12$). For both the distributions, $\mu = U/2$ corresponds to half filling. We take the lattice spacing as the length unit.

We first employ the HF approximation, where the wave function is approximated by a single Slater determinant consisting of a set of orthonormal SP orbitals $\{ \phi_n \}$ ($n$ is an orbital index). The HF equation reads:  
\begin{equation}
\{\mathcal{H}_0+U \sum_{i} ( \langle n_{i \downarrow} \rangle n_{i \uparrow}+ \langle n_{i \uparrow} \rangle n_{i \downarrow} ) \} \phi_n= \epsilon_n \phi_n,
\end{equation}
where $\mathcal{H}_0$ is the one-body part of the Hamiltonian and we neglect $\langle c_{i \uparrow}^\dagger c_{i \downarrow}\rangle$. To find a site-dependent mean-field solution $\langle n_{i \sigma} \rangle$ for the HF equations, we employ the iterative scheme. One typically needs from several to several tens of initial guesses in obtaining convergent physical quantities such as AF order parameters and DOS.

\begin{figure}
 \centering
 \includegraphics[width=.475\textwidth,clip]{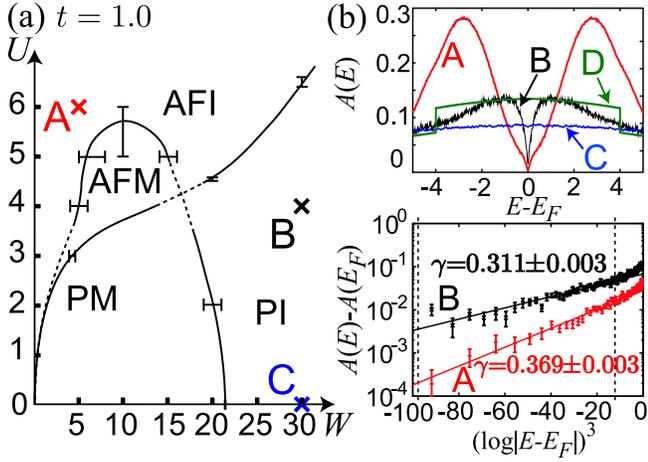}
 \caption{(color online). (a) Ground-state phase diagram of 3D Anderson-Hubbard model at half filling for Gaussian distribution of $P_V$. AFI, AF insulator; AFM, AF metal; PI, paramagnetic insulator (Anderson insulator); PM, paramagnetic metal. (b) DOS with system size $8\times8\times250$: A ($t=1, U=6,W=5$), B ($t=1, U=4,W=30$), C ($t=1, U=0,W=30$), D ($t=0, U=4,W=30$). We employ Lorentz broadening with a broadening factor $1.25\times10^{-3}$ and $6.25\times10^{-4}$ for A and B, respectively. The broken lines denote $|E-E_F|=10^{-2}$ and $10^{-1}$. The DOS fits well with $A(E) \propto\exp(-(-\gamma\log| E-E_F|)^3 )$ shown by the fitting lines for $10^{-2}<|E-E_F|<10^{-1}$ as shown in the lower panel.} 
 \label{fig:pd}
 \vspace{-1em}
\end{figure}
\begin{figure}
 \centering
 \includegraphics[width=.475\textwidth,clip]{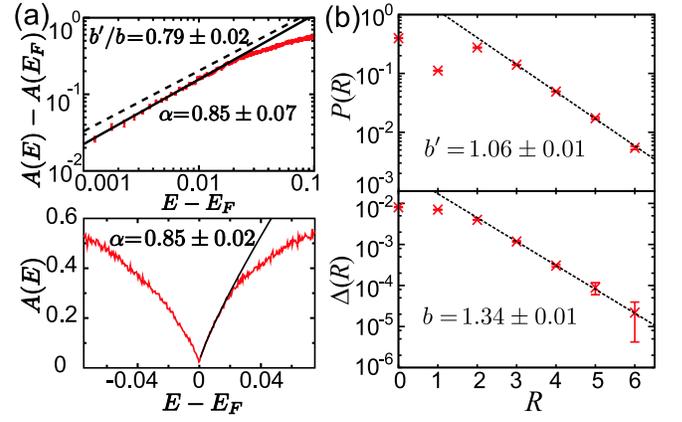}
 \caption{(color online). (a) DOS by HF in 1D at $t=0.3$, $U=1.0$, $W=2.0$, $E_F=U/2-1$ ($\Ns=14$). Fitting of DOS gives $\alpha=0.85 \pm 0.07$ (solid line), which is in good agreement with the expected exponent of $b^\prime/b=0.79\pm0.02$ (broken line). (b) Numerical estimates of $P(R)$ and $\Delta (R)$. Fitting by Eqs.~(\ref{eq:D-R}) and (\ref{eq:P(R)}) gives $b^\prime = 1.06 \pm 0.01$ and $b=1.34 \pm 0.01$.}
 \label{fig:dos-HF}
 \vspace{-1em}
\end{figure}
\begin{figure}
 \centering
 \includegraphics[width=.475\textwidth,clip]{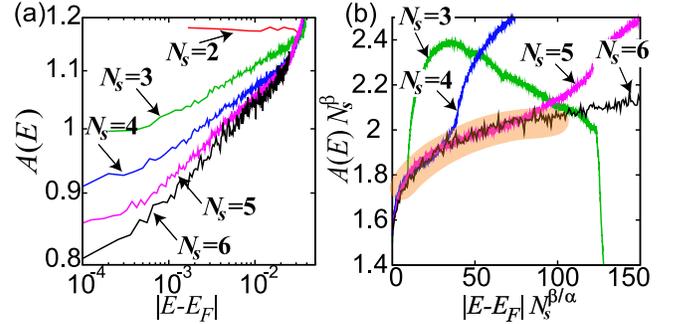}
 \caption{(color online). (a) DOS in 1D with ED (open boundary condition): $t=0.1$, $U=1.0$, $W=1.0$, $E_F=U/2$, ($\Ns=2,3,4,5,6$). We average the DOS over $3.2\times10^7$ realizations of disorder for $\Ns=6$. (b) Scaling plot by $A(\epsilon, \Ns^{-1}) = \Ns^{-\beta} f( \epsilon \Ns^{\beta/\alpha})$ ($\alpha=0.075, \beta=0.375$). }
 \label{fig:ED}
 \vspace{-1em}
\end{figure}
Figure~\ref{fig:pd}(a) shows the ground-state phase diagram at half filling in 3D. We identify insulating phases by extrapolation of the localization lengths $\xi$ to the bulk limit. The localization length $\xi$ is defined by the asymptotic behavior of SP orbitals near $E_F$ at long distances as $\phi_n \propto \exp(-r/\xi)$, where $r$ is the distance from the center of the orbital. We obtain the AF magnetic transition points by fitting the AF magnetic order parameter with the mean-field critical exponent, $1/2$. Detailed analyses of the phase diagram will be discussed elsewhere. For $U<6$ and with increasing $W$ from $0$, metals appear from AFI as in the 2D result~\cite{Heidarian04}, with further reentrant transition to insulators (AFI or PI). Naively one might expect $A(E_\mathrm{F})>0$ for $W>0$. Figure~\ref{fig:pd}(b) shows DOS for typical parameters. Indeed, there are no soft gaps when $U$ or $t$ is zero. However, we find soft Hubbard gaps over the entire insulating phases in the case of $U>0$ and $t>0$ regardless of the AF magnetic order. Although a power law scaling $A(E)\propto |E-E_F|^{\alpha}$ with exponents $0.5<\alpha<1$ looks fit in the range $|E-E_F| > 0.1$ (not shown) being consistent with the previous HF study~\cite{Fazileh06}, closer look for $|E-E_F|<0.1$ fits better with $A(E) \propto\exp(-(-\gamma\log| E-E_F|)^3 )$ with $\gamma>0$ rather than the power-law scaling.

The unconventional soft gaps exist also in 1D regardless of electron filling. Figure~\ref{fig:dos-HF}(a) shows DOS with the HF approximation for the box distribution of $P_V$. Here, holes are partially doped with the chemical potential $\mu$ being shifted by $-1.0$ from the half filling to increase the average distance between electrons to capture long-range asymptotic behavior easily. In contrast to the 3D case, they fit well with a power law $A(E)\propto |E-E_F|^{\alpha}$ even at low energies. The gaps again vanish with the decreasing energy scale when $t$ or $U$ becomes zero (not shown).

In Fig.~\ref{fig:ED}(a), we further show DOS with ED in 1D. We assume a scaling function that $A(\epsilon, \Ns^{-1}) = \Ns^{-\beta} f( \epsilon \Ns^{\beta/\alpha})=\epsilon^\alpha g( \Ns^{-\beta/\alpha} \epsilon^{-1})$ corresponding to $A(\epsilon, \Ns^{-1}=0) \propto  \epsilon^\alpha$ and $A(\epsilon=0, \Ns^{-1}) \propto \Ns^{-\beta}$ ($\epsilon = | E-E_F|$). As shown in Fig.~\ref{fig:ED}(b), DOS well converges to this scaling function with $\alpha=0.075$ and $\beta=0.375$. Although a possible logarithmic scaling cannot be excluded because of the small system size, the ED results are consistent with the HF results and support a mechanism of the soft gap working beyond the mean-field level. Because the soft gap is restricted to very low energies in our 1D study, further analyses at lower energies are desired in 2D, where only a pseudogap has been found so far~\cite{Chiesa08}.

Now we discuss a possible origin of the soft gap. For simplicity without loss of generality, we restrict ourselves to a SP excitation for the electron side, namely, $E>E_F$. We consider the case of $r_\mathrm{int} \ll \xi$, where $r_\mathrm{int}$ is the range of the interaction in the model. For $t \neq 0$, virtual hopping of electrons generates effective interaction, which exponentially decreases with the mutual distance. This effect is not considered in the ES theory which regards electrons as classical particles. The DOS averaged over the random potential is obtained as
\begin{eqnarray}
 A(E) &=&  {\Bigl\langle{\int}_{-\infty}^{\infty} P_V(V_1)A_1(E,V_1) {\rm d} V_1 \Bigr\rangle}_{\{ V_{\overline{1}} \}}, 
\end{eqnarray}
where the symbol $\{ V_{\overline{1}}\}$ denotes a set of random potentials $V_i$ except for $V_1$. Note that $A_1 (E, V_1)$ is the DOS under the condition of the fixed $V_1$ at the site $1$ and implicitly depends on ${\{ V_{\overline{1}}\}}$. Here we decompose the average over the random potential into the part for $V_1$ as described by $\int P_V(V_1) d V_1$ at fixed configurations of $\{ V_{\overline{1}}\}$ and the subsequent average with respect to $\{V_{\overline{1}}\}$.

\begin{figure}
 \centering
 \includegraphics[width=.475\textwidth,clip]{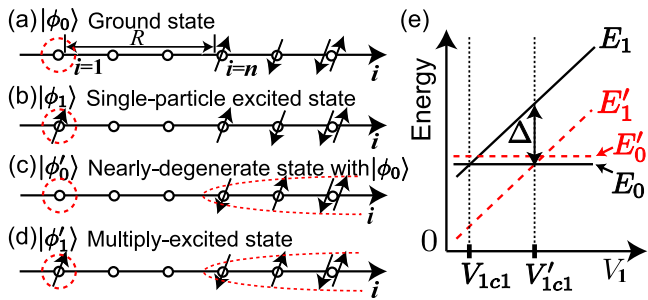}
 \caption{(color online). Schematic illustration of (a) the ground state, (b) a SP excited state, (c) a nearly-degenerate state with the ground state and (d) a multiply-excited state. (e) Schematic of $V_1$ dependence of excitation energies.}
 \label{fig:level-cross}
 \vspace{-1em}
\end{figure}
We discuss $V_1$-dependence of $A_1 (E, V_1)$ for fixed ${\{ V_{\overline{1}}\}}$. When $V_1$ decreases, the ground-state occupation of the site 1 changes from 0 to 1 and then from 1 to 2 at $V_{1c1}$ and $V_{1c2}$, respectively. A possible ground state $|\phi_0\rangle$ at $V_1>V_{1c1}$ is illustrated in Fig.~\ref{fig:level-cross}(a), where the site 1 is empty and the total particle number is $\Ne=N_a$ and the energy $E_{0}(V_1)$. Near $V_{1c1}$ but for $V>V_{1c1}$, a SP excited state $|\phi_1\rangle$ with $\Ne=N_a+1$ and the energy $E_{1}(V_1)$ is defined by the electron configuration except for the site 1 is fixed to be the same as $|\phi_0\rangle$, as is illustrated in Fig.~\ref{fig:level-cross}(b). One might think that $|\phi_1\rangle$ becomes the ground state below $V_{1c1}$, where $\Ne=N_a+1$. In this case, however, the SP excitation gap $E_1-E_0$ vanishes at $V_{1c1}$ leading to absence of a gap in the $V_1$-averaged DOS at the site $1$. Thus the numerical evidences of the soft gaps indicate that $|\phi_1\rangle$ as a SP excited state is excluded by the electron correlation. 

In contrast to the ES theory,  we assume a multi-valley energy landscape, which may be characteristic to random systems. Then there exist many arbitrarily-low-energy excited states whose configurations are the same with $|\phi_0\rangle$ at the site $1$ but globally different on other sites. In Fig.~\ref{fig:level-cross}(c), we illustrate a state $|\phi_0^\prime\rangle$ at a local minimum $E_0^\prime$ nearly degenerate with $|\phi_0\rangle$, whose configurations not only at the occupied site $n$ nearest to the site $1$ at the distance $R$ but also farther sites ($>R$) are relaxed one after another. Figure~\ref{fig:level-cross}(d) shows a SP excited state $|\phi_1^\prime \rangle$ from  $|\phi_0^\prime\rangle$ with the energy $E_{1}^\prime$ and the site-$1$ occupancy identical with $|\phi_1 \rangle$. Here, the two nearly-degenerate states, $|\phi_1\rangle$ and $|\phi_1^\prime\rangle$ are separated by a barrier, where multi-particle relaxation is required to reach from one to the other. Now $E_1$ is given by $(V_1-E_F) + \sum_i U_{1 i}+E_0$, where $U_{1 i}$ is the interaction energy between electrons on the site $1$ and those on the site $i$. Note that only the particles at the sites $i$ which satisfy $R \le |i-1| \lesssim R+\xi$ interact with the site $1$ with the amplitude $|U_{1 i}|$ comparable to $|U_{1 n}|$ because of the localized nature. On the other hand, because $E_0 \simeq E_0^\prime$ and the configurations of $|\phi_1^\prime\rangle$ on these sites are different from those of $|\phi_1\rangle$, $E_1^\prime$ is different from $E_1$ by typically as much as $|U_{1n}|$. Thus one can find $|\phi_1^\prime\rangle$ with the energy $E_1^\prime$ lower than $E_1$ by as much as $|U_{1 n}|$ among many nearly-degenerate states with $|\phi_1\rangle$. Now $E_1^\prime(V_1)$ and $E_0(V_1)$ crosses at $V_1=V_{1c1}^\prime$ and for $V_1<V_{1c1}^\prime$ the ground state becomes $|\phi^\prime_1\rangle$.  Note that the excitation energy $E_1^\prime-E_1$ is negative very near $V_{1c1}^\prime$ but for $V_1>V_{1c1}^\prime$. The state $|\phi^\prime_1\rangle$ is not counted in DOS, because this state is not a SP excitation of $|\phi_0\rangle$, but rather a multiply-excited state. Thus the energy difference $\Delta = |E_1(V_{1c1}^\prime) - E_1^\prime(V_{1c1}^\prime) |$ is the lowest energy of SP excitations counted in $A$ near $V_1=V_{1c1}^\prime$.

One might think that, as in the ES theory, it is possible to lower the energy of $|\phi_1 \rangle$ from $E_1$ to $E_1^\prime$ by relaxing \textit{local} electronic configurations only near the site $n$. It, however, always increases the energy of the electrons other than those on the site $1$, because they have already been optimized in the ground state and the increase dominates at large $R$. Thus a global reconstruction is required to lower the energy.

From the above discussion, $\Delta$ scales as
\vspace{-0.3em}
\begin{eqnarray}
 \Delta(R) &=&  a\exp(-b R), \label{eq:D-R}
\end{eqnarray}
\vspace{-0.3em}
where $a$ and $b$ are non-universal positive constants. Hereafter we neglect logarithmic corrections. Under the assumption of linear dependence of the excitation energies on $V_1$ as shown in Fig.~\ref{fig:level-cross}(d), the local DOS averaged by $V_1$ has a gap of $\Delta$ as follows;
\begin{eqnarray}
 {\small \int}_{V_{1c1}^\prime}^{\infty} P_V(V_1)A_1(E,V_1) {\rm d} V_1 &\propto& H_\mathrm{s} (E-E_F-\Delta), \label{eq:local_A}
\end{eqnarray}
where $H_\mathrm{s}$ is the Heaviside step function. The same argument applies around $V_1=V_{1c2}$.

The distribution of $R$ with respect to $\{ V_{\overline{1}}\}$ follows 
\vspace{-0.3em}
\begin{eqnarray}
 P(R) &=&  a^\prime \exp(-b^\prime R^d), \label{eq:P(R)}
\end{eqnarray}
\vspace{-0.3em}
at long distances, where $a^\prime$, $b^\prime$ are non-universal positive constants again. Equations~(\ref{eq:D-R}) and (\ref{eq:P(R)}) lead to
\vspace{-0.3em}
\begin{equation}
 Q(\Delta)=P(R(\Delta)) \left| \frac{dR}{d\Delta} \right| \propto \Delta^{-1} \exp({-\frac{b^\prime}{b^d} (-\log \Delta)^d}), \label{eq:P(D)}
\vspace{-0.3em}
\end{equation}
where $Q(\Delta)$ are the distribution function of $\Delta$. Equations~(\ref{eq:local_A}) and (\ref{eq:P(D)}) lead to
\begin{equation}
 A(E)\propto{\small \int}^{|E-E_\mathrm{F}|}_{0} {\rm d} \Delta Q (\Delta) \propto \exp({-\frac{b^\prime}{b^d} (-\log |E-E_F|)^d}), \label{eq:DOS-HF}
\end{equation}
which is consistent with the observed scaling in 1D and 3D, further also in 2D (not shown) within the HF approximation. We also confirmed that this scaling is equally valid for a discrete distribution of $P_V$ (not shown). For $d=1$, this leads to a power law with a non-universal exponent: $A(E) \propto {|E-E_\mathrm{F}|}^{b^\prime/b}$. Non-universal power-law distributions of energies without divergence of any length scales are common in Griffith phases~\cite{Griffith69}. Equation~(\ref{eq:D-R}) indicates that $a$, namely the energy scale of the gaps vanishes as $t$ or $U$ vanishes. Furthermore, the exponent $\alpha=b^\prime/b$ is expected to decrease as $t$ becomes smaller because of the reduction of $\xi$. These predictions are consistent with our HF results in 1D. However, it conflicts with a DMFT study~\cite{Dobrosavljevic97} and some mean-field studies~\cite{Dobrosavljevic03, Byczuk05} which exhibit absence of the soft gaps. This may be because they ignore spatial correlations. The latter ignore inhomogeneity of the electronic structures. Indeed, a DMFT study with the intersite self-energy retrieves the suppression of DOS near $E_F$~\cite{Song08}.

In Fig.~\ref{fig:dos-HF}(b), we show a further numerical evidence of our theory in 1D. Figure \ref{fig:dos-HF}(b) shows $\Delta(R)$ and $P(R)$ calculated by the following procedure: First, we obtain the ground state for each realization of random potentials. We construct the lowest SP excited state by adding one electron to the lowest unoccupied orbital. Next we optimize the mean fields by the iterative scheme starting from those of the SP excited state with $\Ne$ fixed. Then $\Delta$ is obtained as the difference of these two excitation energies. We calculate $R$ as the distance between the center of the lowest unoccupied orbital, $r$ and those of the occupied orbitals nearest to $r$ in the ground state. We define the center of the orbital as the site which has the maximum weight. Fitting by Eqs.~(\ref{eq:D-R}) and (\ref{eq:P(R)}) gives $b^\prime = 1.06 \pm 0.01$, $b=1.34 \pm 0.01$. Estimated exponent of $b^\prime/b=0.79\pm0.02$ is in good agreement with $\alpha=0.85 \pm 0.07$ obtained directly from DOS. This is a numerical evidence for the validity of our theory. 

Although the power-law was proposed to interpret the photoemission experiments~\cite{Kim06, Maiti07}, our HF results in 3D indicates that a different asymptotic behavior of the soft Hubbard gap emerges at lower energies, namely, $<10$ meV. Since recent development of photoemission spectroscopy now allows us high-resolution measurement down to $1$ meV, we believe that our paper provides incentive for such high-resolution photoemission experiments as well as for other measurement such as electrical transport measurement near the MITs.

In summary, we have found an unconventional type of soft gaps in the Anderson-Hubbard model, although only short-range interaction is present. To clarify their possible origin, we have constructed a phenomenological theory. Detailed comparisons between our theory and the non-ES soft gaps observed in experiments are left as a future challenge.

We thank B. Shklovskii for fruitful discussions. M. I. thanks Aspen Center for Physics for the hospitality. Numerical calculation was partly carried out at the Supercomputer Center, Institute for Solid State Physics, Univ. of Tokyo. This work is financially supported by MEXT under the grant numbers 16076212, 17071003 and 17064004. H. S. thanks JSPS for the financial support. 

\end{document}